\documentclass{iucrjournals}
\usepackage{siunitx}
\usepackage{subcaption}
\usepackage{amsmath}
\usepackage{float} 
\DeclareSIUnit{\photon}{photon}
\usepackage{xcolor}

\title{Multi-modal strain mapping of steel crack tips with micrometer spatial resolution}

\author[a,b]{Ahmar Khaliq\IUCrAufn{ahmar.khaliq@uni-siegen.de}}%
\author[a,b]{Felix Wittwer}%
\author[c]{Anna Wildeis}
\author[c]{Markus Hartmann}
\author[c]{Matthias Thimm}
\author[c]{Robert Brandt}
\author[d]{Dennis Brueckner}
\author[d]{Jan Garrevoet}
\author[d]{Gerald Falkenberg}
\author[a,b]{Peter Modregger}%
\affil[a]{Physics Department, University of Siegen, 57072 Siegen, Germany}
\affil[b]{Center for X-ray and Nano Science CXNS,
Deutsches Elektronen-Synchrotron DESY, 22607 Hamburg, Germany}
\affil[c]{Mechanical Engineering Department,
University of Siegen, 57076 Siegen, Germany}
\affil[d]{Deutsches Elektronen-Synchrotron DESY, 22607 Hamburg, Germany}

\begin{document} 
\maketitle 
%\input{iucr.bbl}
%\begin{center}
%\textit{Corresponding author: ahmar.khaliq@desy.de}
%\end{center}

\newcommand{\averagestrain}{\varepsilon_{\mathrm{avg}}}

\begin{abstract}

Due to their superior fatigue strength, martensitic steels are the material of choice for high cyclic loading applications such as coil springs. However, crack propagation is influenced by residual stresses and their interaction is poorly understood. In fact, Linear Elastic Fracture Mechanics predicts un-physical singularities in the strain around the crack tip. In this study, we have combined synchrotron-based x-ray diffraction, x-ray fluorescence, and optical microscopy to map the factual strain fields around crack tips with micrometer spatial resolution. X-ray fluorescence and optical images were co-registered to locate the crack in the x-ray diffraction maps. Observed crystal recovery close to cracks confirmed that the diffraction signal originates at least in parts from the cracks. The retrieved local strain field around the crack was further improved by averaging information over carefully selected diffraction peaks. This procedure provided strain maps around crack tips with a spatial resolution of about~\qty{1}{\micro\meter} and enabled a prediction of further crack growth.
\end{abstract} 


\section{Introduction}

\begin{figure}[htbp]
\centering
\includegraphics[width=0.7\linewidth]{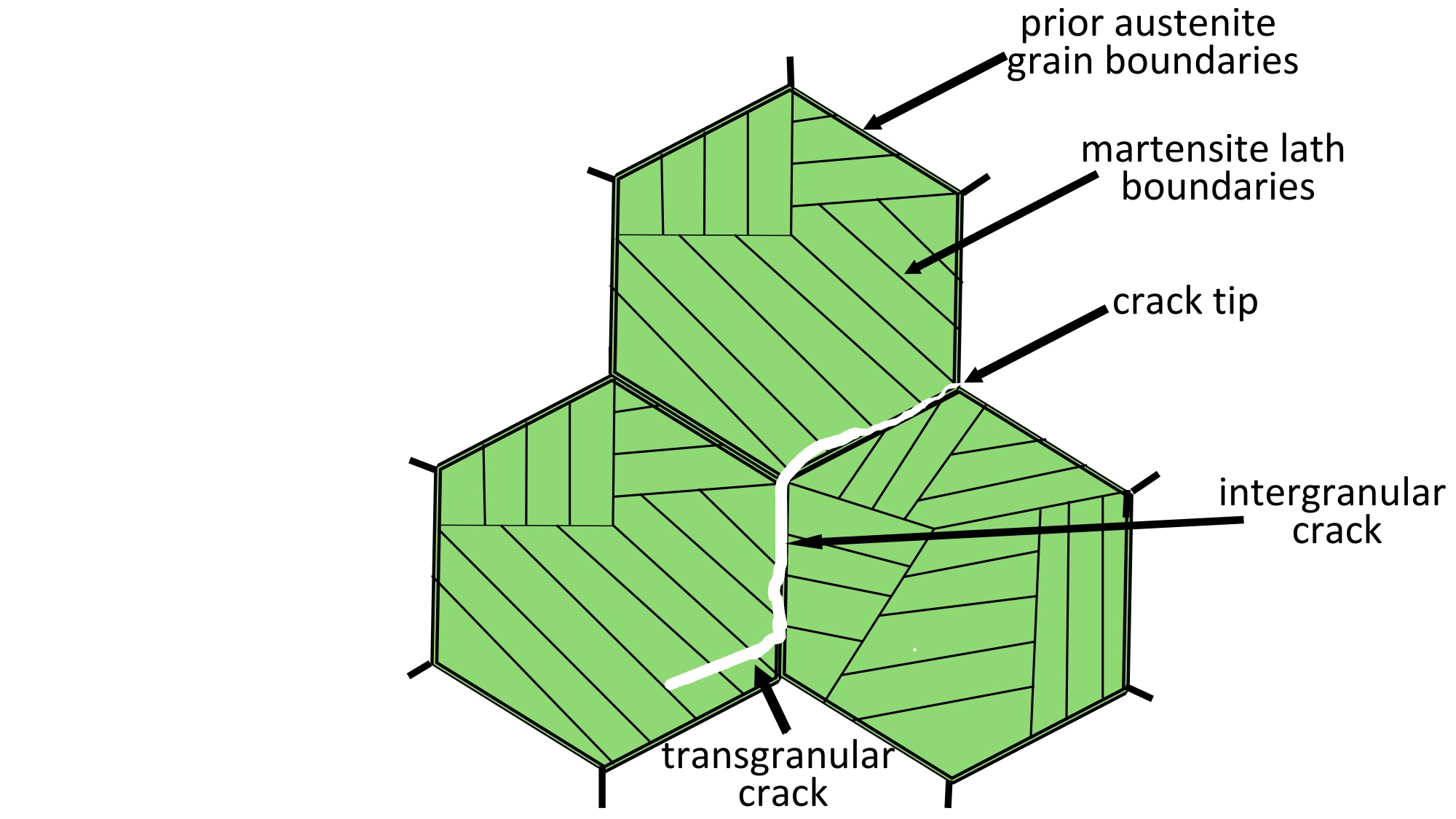}
\caption{Schematic representation of crack propagation in martensite.}
\label{fig:cracksketch}
\end{figure}

Martensitic steel is widely used in engineering applications due to its high strength, high wear resistance~\cite{wu2023enhanced}, and favorable fatigue properties~\cite{heshmati2023correlation, sun2020effects, zhou2024investigation}. The transformation from austenite to martensite occurs through a diffusionless process~\cite{shewmon1969transformations, kelly2012crystallography}, resulting in a highly distorted lattice structure that enhances its mechanical performance~\cite{hutchinson2022origins}. Resulting residual stresses and microstructural heterogeneities play a crucial role in mechanical behavior of martensitic steels, influencing fatigue crack initiation, which can occur intergranularly along grain boundaries and transgranularly through the grains~\cite{wildeis2022influence, kim2022microstructure, park2024effect}. The propagation of short cracks is influenced by multiple microstructural factors such as the prior austenite grain boundaries~\cite{kunio1979role}, lath morphology~\cite{sun2025morphology} or dislocation density~\cite{saha2020martensite}. These factors are also depicted in Fig.~\ref{fig:cracksketch}. 

Linear Elastic Fracture Mechanics (LEFM) provides a framework for analyzing crack behavior in materials, assuming linear elasticity~\cite{anderson2005fracture, broberg1999cracks}. According to LEFM, the stress \(\sigma(r)\) near a crack tip is given by:
\begin{equation}
    \sigma(r) = \frac{k}{\sqrt{2\pi r}},
\end{equation}
where $k$ is a stress intensity factor and $r$ is the distance from the crack tip. At $r=0$ this equation predicts a singularity (i.e., infinite stress), which is not physically. A similar situation holds true for the strain around a crack tip.

The simultaneously acquisition of synchrotron-based x-ray diffraction (XRD) and x-ray fluorescence (XRF) has been proven to be a powerful tool for enhancing the understanding of material properties on the micrometer scale. Examples include the interplay of local strain and material composition on the performance of thin film solar cells~\cite{Ulvestad2019,Calvo-Almazan2019}, the characterization of extraterrestrial samples~\cite{Lanzirotti2024} or the application in chemical analysis~\cite{Su2024}. In this study, we combined micro XRD, XRF, and optical microscopy to map strain fields around crack tips with micrometer resolution in order to determine the factual strain around crack tips. This was done for three samples, two shot-peened samples and one un-modified sample. 

\section{Materials and Methods}

The following section details the methods employed in this study, covering sample preparation, optical microscopy for crack monitoring, and synchrotron XRD and XRF experiments at PETRA~III, DESY.

\subsection{Sample preparation}

The specimens were prepared from high-strength martensitic spring steel SAE 9254 (DIN/EN: 54SiCr6). The steel was austenitized under vacuum at 1080$^{\circ}$C for 100 minutes, quenched with nitrogen, and tempered at 400$^{\circ}$C for 1 hour in argon to form a martensitic structure. Following heat treatment, three samples were machined from 12~mm diameter wire rods via electric discharge machining, featuring a gauge section of 10~mm length, 5~mm width, and 2~mm thickness to promote surface crack initiation. Two specimens were shot-peened~\cite{guagliano2001relating} with an Almen intensity of 0.16~mm using a pneumatic system equipped with dual nozzles, operated at 1.5~bar pressure, and steel shots with a diameter of 0.4~mm (700 HV, G3 per VDFI 8001). This ensured full surface coverage and introducing compressive residual stresses estimated at 900~MPa~\cite{wildeis2021characterizing, wildeis2022influence}. The third specimen remained unpeened.
% PM: Prerana should have cite for shot-peening

To study crack behavior, cracks were initiated by applying uniaxial cyclic loading at a stress ratio of $R = -1$ (i.e., the ratio of minimum to maximum stress), a frequency of 10~Hz, and stress amplitudes varied between 550~MPa and 680~MPa, corresponding to high cycle fatigue conditions. Slip bands formed at prior austenite grain boundaries, acting as crack initiation sites, with shot-peened specimens showing delayed crack growth due to residual stresses~\cite{wildeis2021characterizing, wildeis2022influence}.

\subsection{Optical microscopy}

Optical microscopy was used to track crack initiation and early propagation across the specimens. Surface preparation involved progressive grinding with SiC paper up to grit 4000, followed by a final polish with a colloidal silicon suspension of \qty{0.25}{\micro\meter} grain size to yield a smooth, reflective finish. A confocal laser microscope (Olympus LEXT OLS4000) was employed to acquire detailed images of the specimen surfaces at scheduled intervals during fatigue testing. These images allowed for the measurement of crack lengths and the evaluation of crack density, offering insights into the progression of fatigue damage across both treated and untreated conditions~\cite{wildeis2021characterizing,wildeis2022influence}.

\subsection{Synchrotron-radiation experiment}

\begin{figure}[htbp]
    \centering
    \includegraphics[width=1\textwidth]{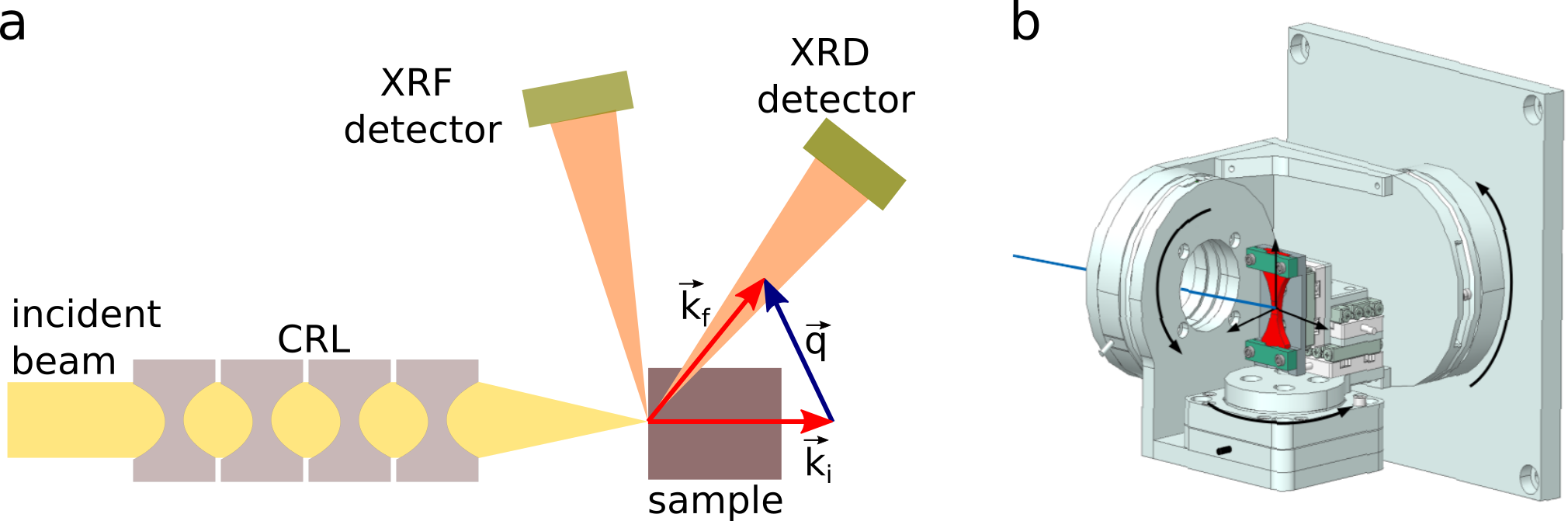}
    \caption{(a) Schematic of the experimental setup. (b) Goniometer configuration used for sample positioning.}
    \label{fig:setup}
\end{figure}

The synchrotron-radiation experiment was conducted at the P06 beamline of PETRA III at DESY, Hamburg~\cite{Falkenberg2020}. A sketch of the utilized setup is shown in Fig.~\ref{fig:setup}(a), which was also described in~\cite{chakrabarti2022x}. A monochromatic x-ray beam of~\qty{35}{\keV} was selected by a Si (111) double-crystal monochromator. The beam was focused using compound refractive lenses (CRLs), achieving a beam size of $h\times v =~\qtyproduct{0.9 x 0.4}{\micro\meter}$. The sample was mounted on a six-axis goniometer, which allowed precise alignment and positioning during measurements (see Fig.~\ref{fig:setup}(b)). The XRD signal was recorded using a detector with a~\qty{55}{\micro\meter} pixel size, positioned approximately~\qty{1.0}{\meter} downstream of the sample and horizontally inclined at~\ang{25} relative to the incident beam. Simultaneously, XRF data was collected using a silicon drift detector (Hitachi High-Tech), positioned slightly offset from~\ang{90} relative to the incident beam for effective fluorescence detection. Two scans were performed for each sample: one overview scan, covering the entire crack with a step size of~\qty{10}{\micro\meter} and an exposure time of~\qty{0.2}{\second} per point, and another focused on the crack tip with a step size of~\qty{1}{\micro\meter} and the same exposure time of~\qty{0.2}{\second}. 

\section{Data Analysis}

In this section, we focus on processing XRD and XRF data through key steps: collecting diffraction patterns, fitting peaks, locating cracks, analyzing effects of crystal recovery, quantifying strain, and mapping the strain field at the crack-tip in martensitic steel samples.

\begin{figure}[htbp]
    \centering
    \includegraphics[width=1\textwidth]{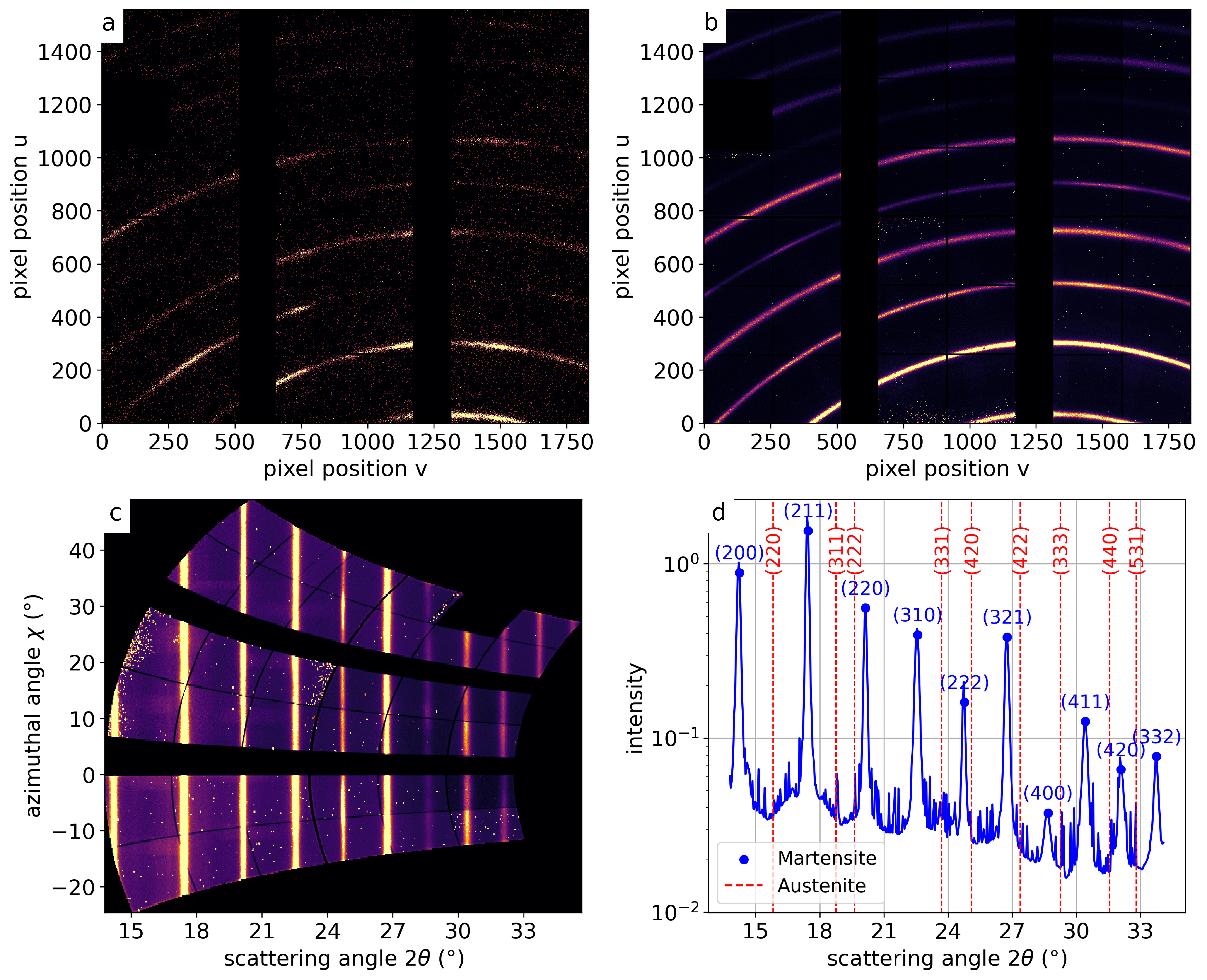}
    \caption{(a) A diffraction pattern of a single scan point of a martensitic steel sample. (b) Diffraction pattern summed over all 14091 scan points. (c) Transformation to polar coordinates of the summed diffraction pattern via caking. (d) Resulting 1D integration with the $2\theta$ positions of martensite and austenite indicated.}
    \label{fig:diffraction_processing}
\end{figure}

The experimental geometry was calibrated using the \texttt{pyFAI} calibration routine~\cite{Ashiotis:fv5028} with Lanthanum hexaboride (LaB\textsubscript{6}) as a diffraction standard. For automatic peak detection diffraction patterns from all scan points (see Fig.~\ref{fig:diffraction_processing}(a)) were summed (see Fig.~\ref{fig:diffraction_processing}(b)). Individual diffraction patterns were transformed from cartesian coordinates (u, v) to polar coordinates ($2\theta, \chi$) via a caking process~\cite{kieffer2013pyfai}, with an example shown in Fig.~\ref{fig:diffraction_processing}(c). Subsequent azimuthal integration yielded the diffraction signal as a function of $2\theta$. As shown in Fig.~\ref{fig:diffraction_processing}(d), the first 10 martensitic diffraction peaks, ranging from 200 to 332 were identified.

\begin{figure}[htbp]
    \centering
    \includegraphics[width=1\textwidth]{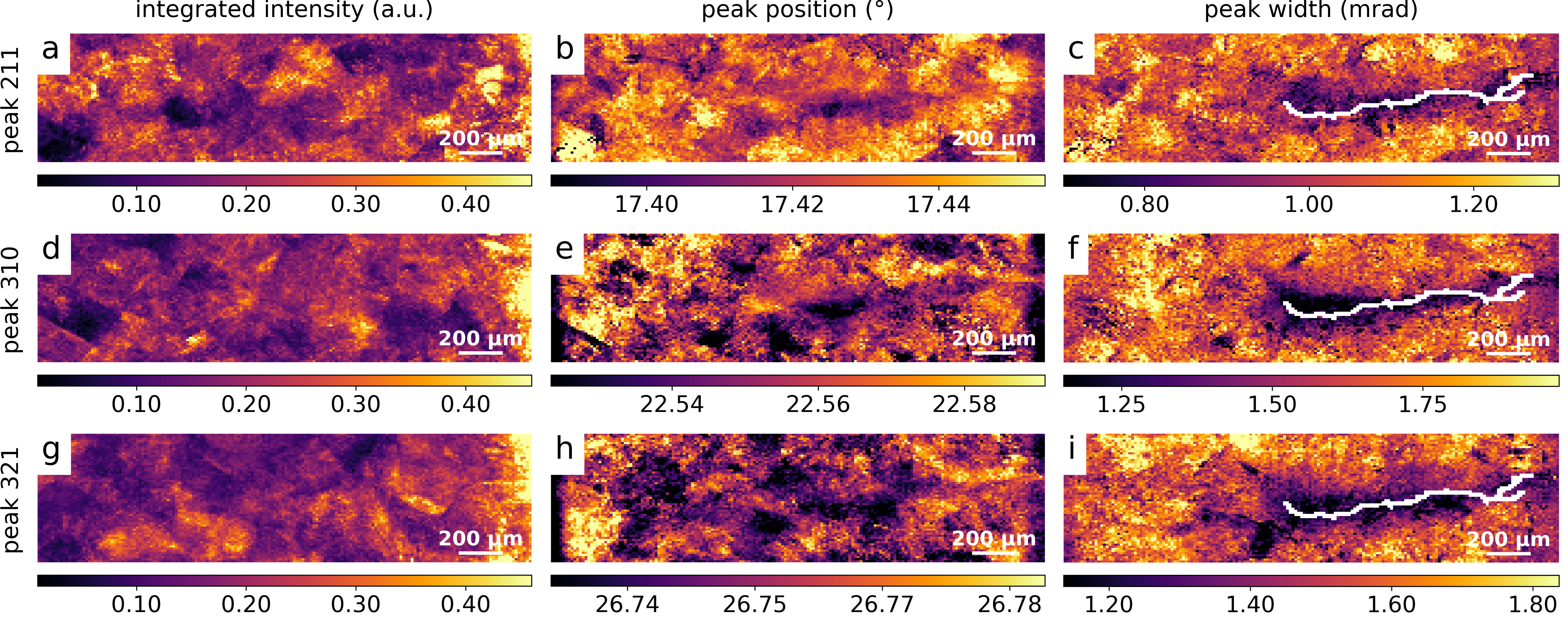}
    \caption{Maps from the overview scan showing integrated intensity (left column), peak position (middle column), and peak width (right column) for peaks 211, 310, and 321. Rows correspond respectively to: peak 211 in (a–c), peak 310 in (d–f), and peak 321 in (g–i).}
    \label{fig:overview_maps_initial}
\end{figure}

For the determination of the peak height H, the angular peak position \(\mu\), the angular peak width \(\sigma\), and the background $C$ of all occurring diffraction peaks, the azimuthal integrated intensities were fitted to a Gaussian distribution given by
\begin{equation}
    g(2\theta) = H \exp \left( -\frac{1}{2} \left(\frac{2\theta - \mu}{\sigma} \right)^2 \right) + C.
\end{equation}
To study local variations, using the fitted Gaussian parameters, maps for peak height, peak position, and peak width were computed for each scan point of each peak. The integrated intensity \(I_{int}\) was obtained using \(I_{int} = H \cdot \sqrt{2\pi} \cdot \sigma\). Maps of the integrated intensity, peak position and peak width for the overview scan were obtained for all peaks; maps for peaks 211, 310, and 321 are shown as examples in Figs.~\ref{fig:overview_maps_initial}(a-i). The lattice parameter of unstrained martensitic steel was estimated by the observed, averaged lattice parameter $a_0$. The latter was retrieved from 10 peaks in the diffraction pattern averaged over an entire scan. The result was $a_0 = 2.866(4) \, \text{\AA}$, which is in agreement with published values~\cite{Xiao1995}.

\begin{figure}[htbp]
    \centering
    \includegraphics[width=1\textwidth]{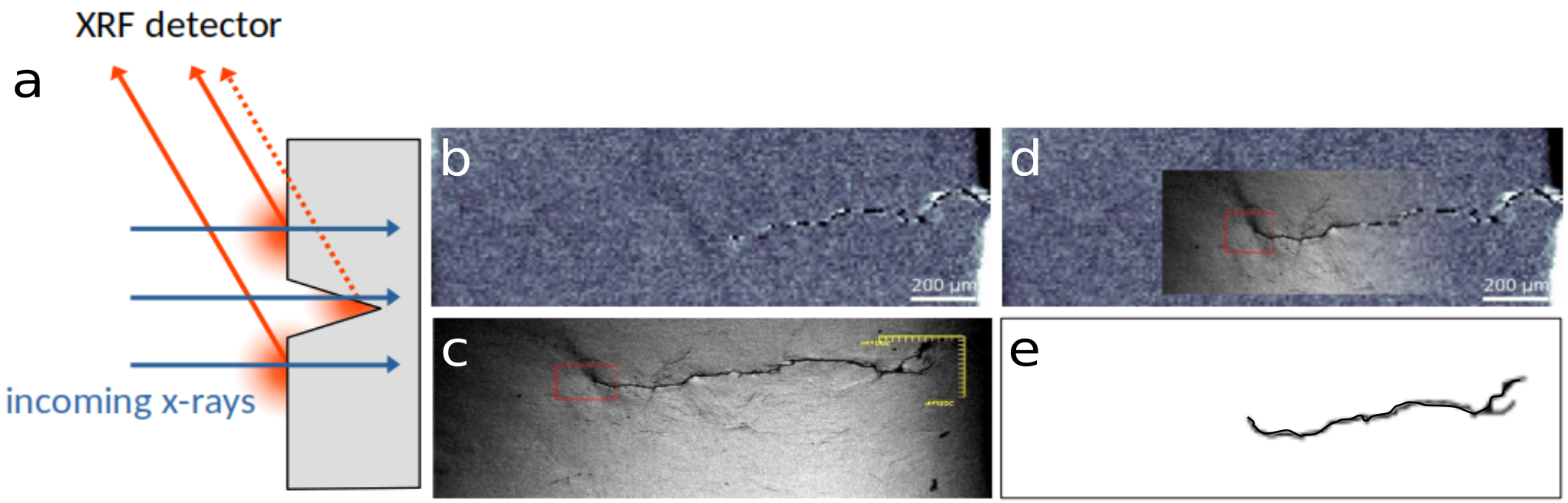}
    \caption{Process of crack localization using XRF and microscopy: (a) Schematic of the XRF process, (b) XRF image, (c) Optical microscope image, (d) Registered XRF and microscopy images, (e) Crack segmentation.}
    \label{fig:crack_registration}
\end{figure}

Since cracks are not directly visible in the XRD maps, the cracks were located using the self-absorption of fluorescent x-rays within the material, which blocks the XRF signals and creates highlighted and shadowed areas in the XRF image (similar to scanning electron microscopy~\cite{baba1994shadow}), as shown in Fig.~\ref{fig:crack_registration}(a). The resulting XRF image is shown in Fig.~\ref{fig:crack_registration}(b). However, due to the limited spatial resolution and contrast, crack tip segmentation directly from XRF data was challenging. To achieve accurate crack segmentation, optical microscope images (see Fig.~\ref{fig:crack_registration}(c)) were acquired and registered onto the summed XRF image (total XRF), enabling crack localization, as depicted in Fig.~\ref{fig:crack_registration}(d). The resulting crack silhouette is shown in Fig.~\ref{fig:crack_registration}(e).

\begin{figure}[htbp]
    \centering
    \includegraphics[width=1\textwidth]{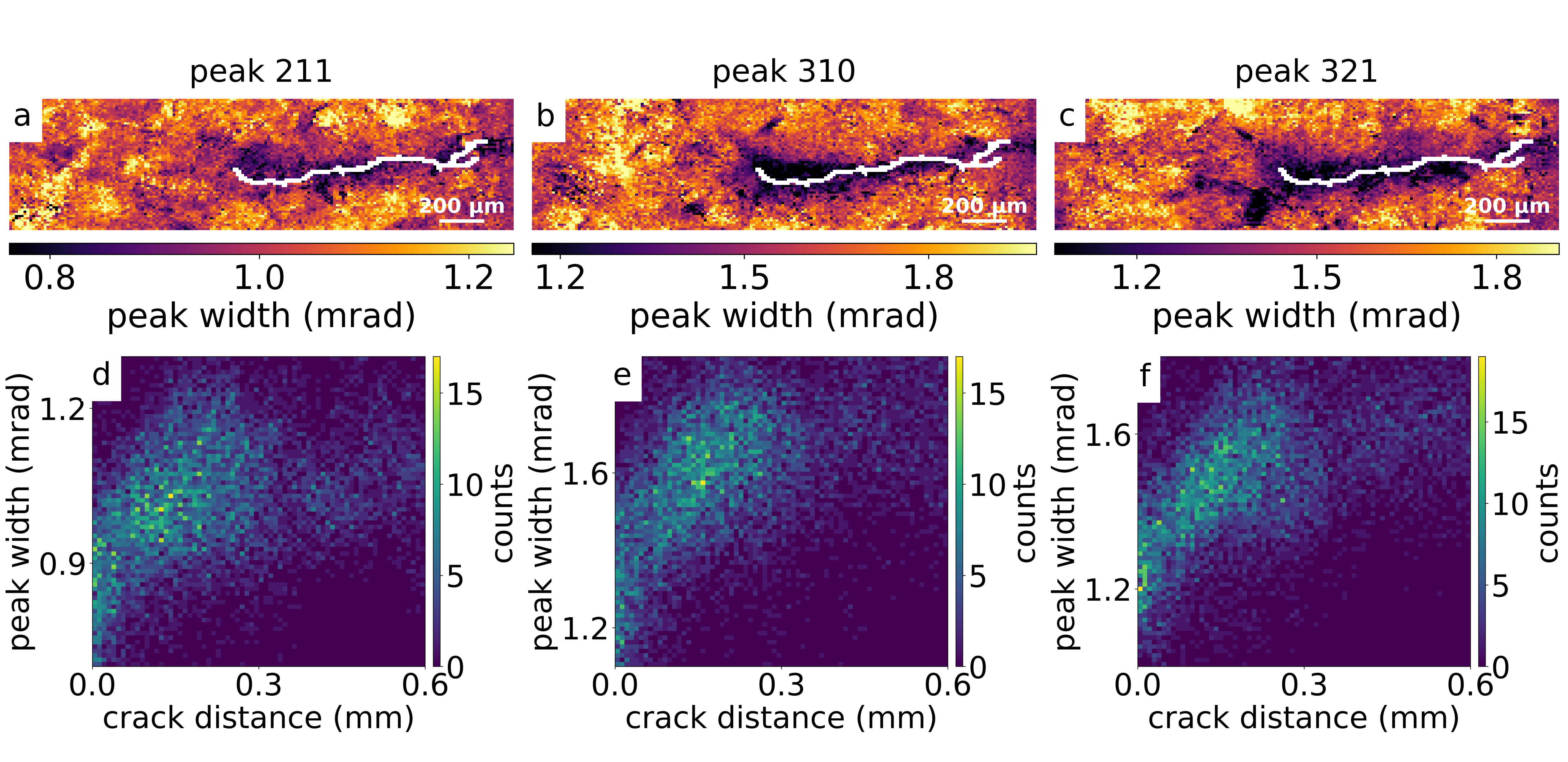}
    \caption{Peak width maps with crack overlay and their corresponding 2D histograms of peak width vs crack distance: (a+d), (b+e), and (c+f). Here, we observe effects of crystal recovery up to~\qty{300}{\micro\meter} from the crack for the peaks 211, 310, and 321, respectively.}
    \label{fig:recrystallization_analysis}
\end{figure}

The crack silhouette (see Fig.~\ref{fig:crack_registration}(e)) was overlaid onto the XRD maps to investigate variations near the crack. As an example, three peak width maps with the crack overlay are shown in Figs.~\ref{fig:recrystallization_analysis}(a-c). 
All three maps show a decrease in peak width close to the crack. This is verified by the 2D histogram of the peak widths versus crack distance, which show a trend up to~\qty{300}{\micro\meter} from the crack, see Figs.~\ref{fig:recrystallization_analysis}(d-f). According to the Scherrer equation~\cite{williamson1953x}, this indicates an increased crystallite size close to the crack. This change in crystallite size could be caused by recrystallization, however previous measurements showed no evidence of recrystallization in the vicinity of the crack~\cite{wang2024unraveling, wildeis2022influence}. Instead, it is likely caused by dynamic recovery which reduces local microstrain and dislocation density and thereby decreases peak width~\cite{wang2024unraveling, miao2017evaluation}. Furthermore, this confirms that at least some diffraction information originated from the vicinity of the crack.

The strain along the scattering vector \(\varepsilon_q\) was determined using the differential Bragg equation~\cite{hart1969high}:
\begin{equation}
    \varepsilon_q = -\frac{\Delta\theta}{\tan\theta_0},
\end{equation}
where \(\theta_0\) is the average peak position (corresponding to $a_0$), and \(\Delta\theta = \theta_{\text{measured}} - \theta_0\) represents the peak shift at each scan point.

\begin{figure}[htbp]
    \centering
    \includegraphics[width=0.6\textwidth]{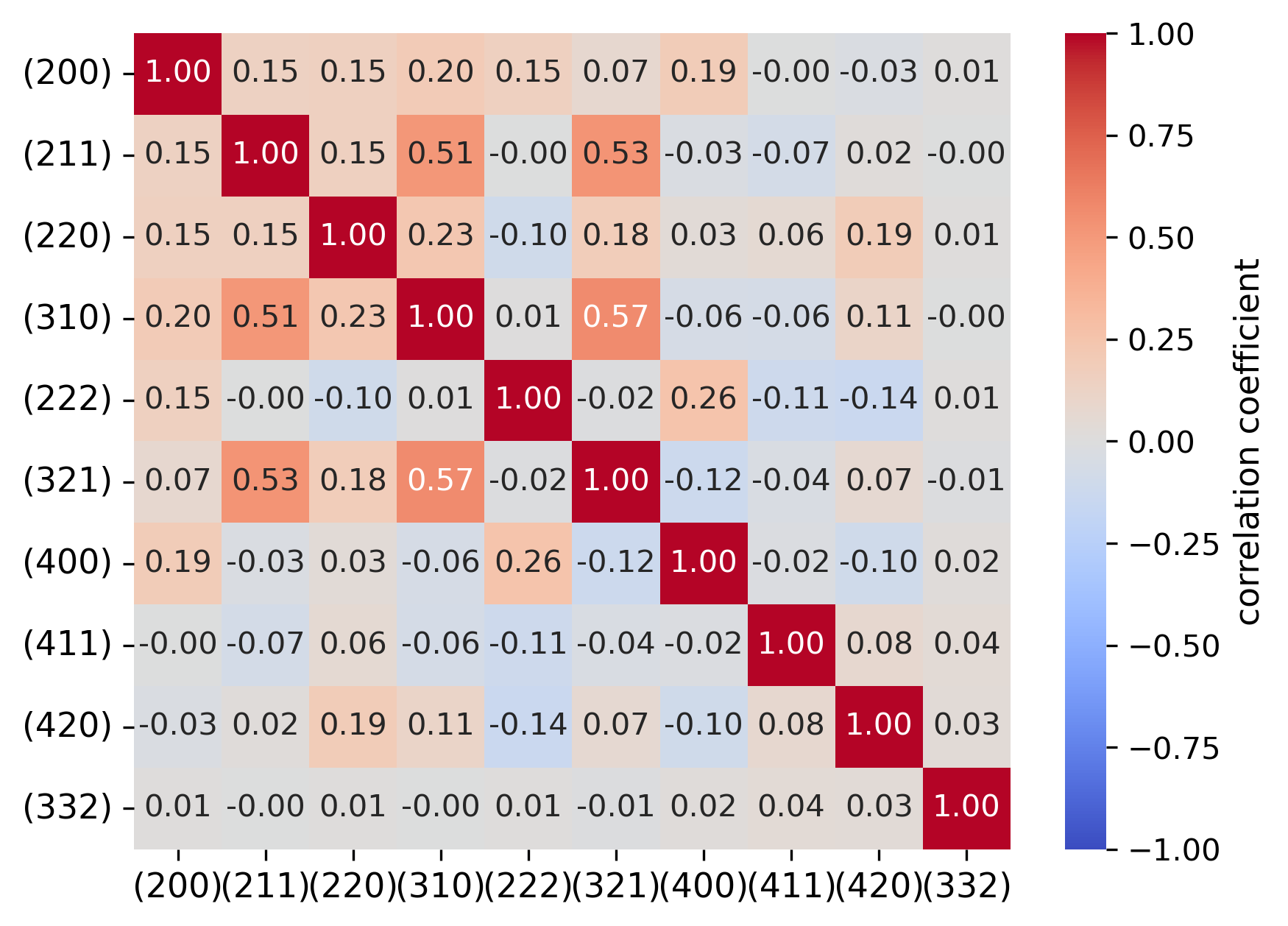}
    \caption{Heatmap of correlation coefficients of peak width maps, which we used to identify the following highly correlated peaks: 211, 310, and 321.}
    \label{fig:correlation_heatmap}
\end{figure}

To improve the contribution from the crack vicinity, we combined the strain information from several different diffraction peaks. To this end, we have calculated the pair-wise correlation coefficient between peak width maps of different peaks. The corresponding heatmap is shown in Fig.~\ref{fig:correlation_heatmap}. Here, we have selected three peaks with the triple (i.e., 211, 310 and 321) of largest correlation coefficients ($r=0.51-0.57$), which are -- not coincidentally-- the peaks with highest multiplicity.

The projection of the two-dimensional strain tensor $\varepsilon_{ij}$ in terms of the sample surface coordinate system (see Fig.~\ref{fig:setup}(a), $z$ direction pointing into the sample) can be calculated according to $\varepsilon_q = \sum_{ij} q_i q_j \varepsilon_{ij}$~\cite{ramirez2016stress} with the direction of the scattering vector and its corresponding components ${\bf q} = (q_1, q_3) = (\cos\theta,-\sin\theta)$. This yields
\begin{equation}
    \varepsilon_q = \varepsilon_{11}\cos^2\theta + \varepsilon_{33}\sin^2\theta - \varepsilon_{13}\sin(2\theta).
\end{equation}
Thus, the average of the strain from the 211, 310, and 321 peaks was
\begin{equation}
    \averagestrain = 0.963 \varepsilon_{11} + 
    0.0373 \varepsilon_{33} -0.374 \varepsilon_{13},
\end{equation}
which is predominately along the surface.

\begin{figure}[htbp]
    \centering
    \includegraphics[width=1\textwidth]{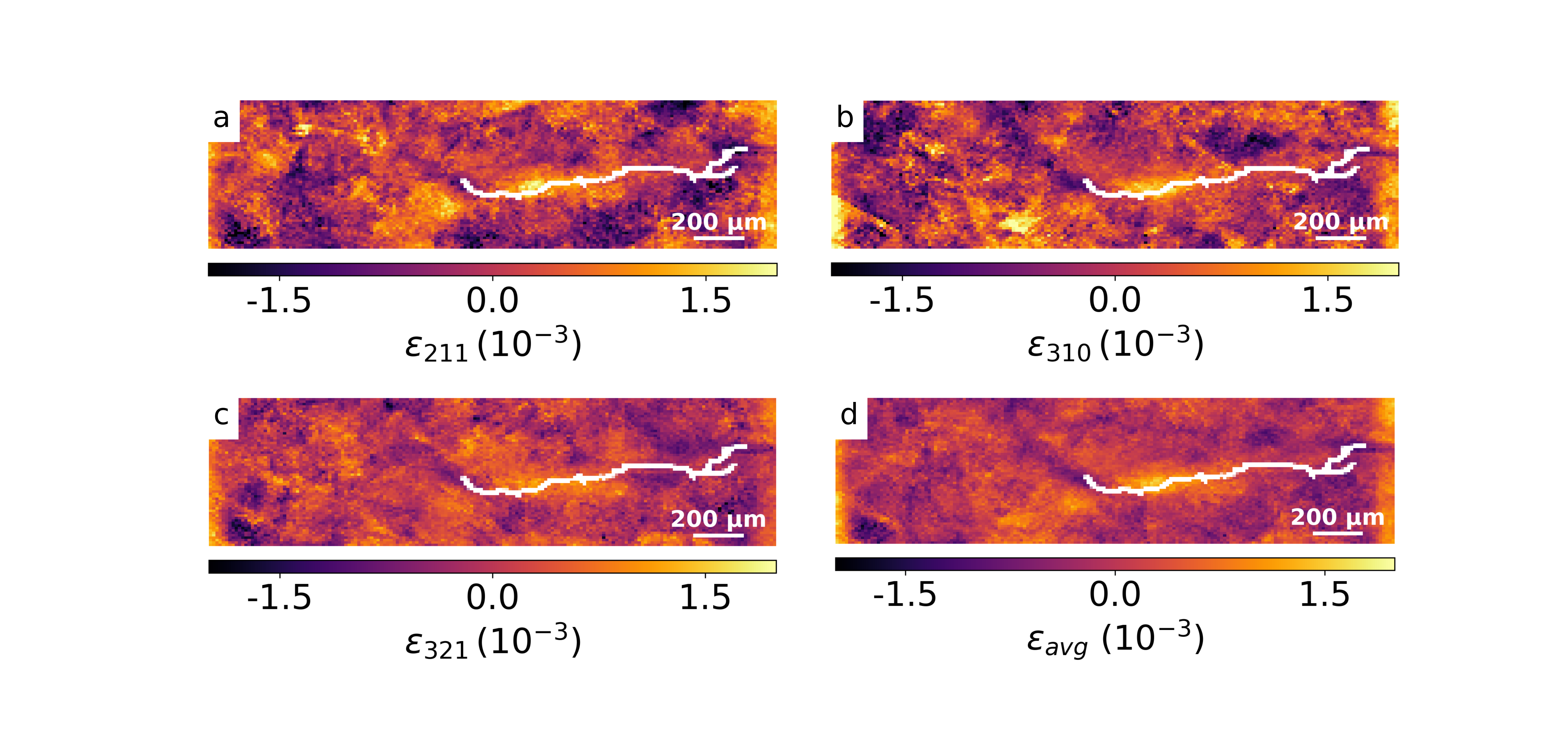}
    \caption{Strain projection maps with crack overlay for peaks (a) 211, (b) 310, (c) 321, and (d) displays the averaged strain map $\averagestrain$ of a shot-peened sample.}
    \label{fig:epsilon}
\end{figure}

The strain projections from the selected peaks were first centered by subtracting their respective mean values and then combined by averaging. The contributing strain projections and the resulting average strain $\averagestrain$ are shown in Fig.~\ref{fig:epsilon}. The approach of combining information from several peaks is somewhat justified by the apparent improved homogeneity in Fig.~\ref{fig:epsilon}(d) compared Figs.~\ref{fig:epsilon}(a-c).

\begin{figure}[htbp]
    \centering
    \includegraphics[width=0.5\linewidth]{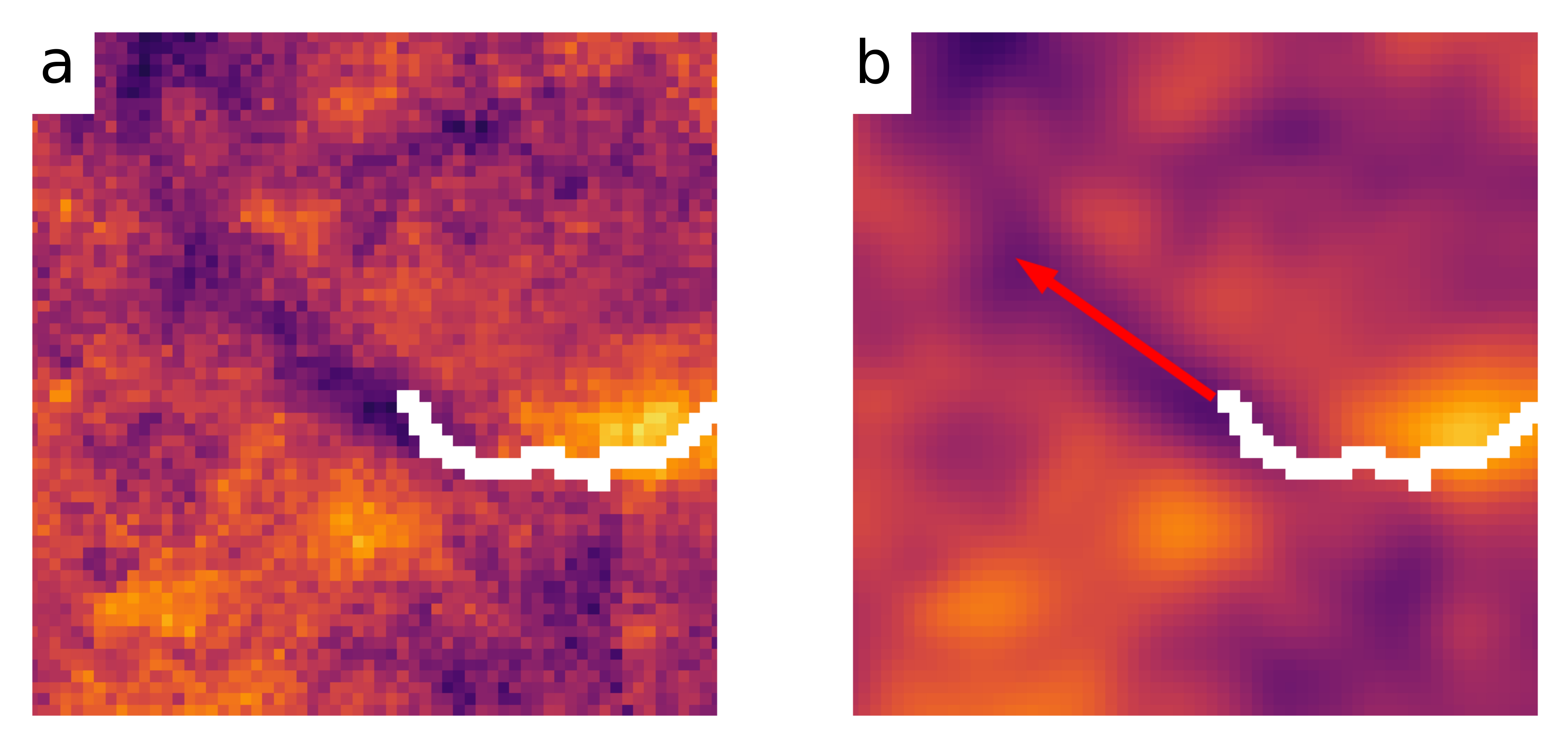}
    \caption{Heuristic prediction of crack propagation. (a) Zoom into figure~\ref{fig:epsilon}d. (b) Predicted path of crack propagation overlaid on a blurred version of (a).}
    \label{fig:crack_prediction}
\end{figure}

Fig.~\ref{fig:crack_prediction}(a) shows a zoom at the crack tip of the retrieved average strain shown in Fig.~\ref{fig:epsilon}(d). A heuristic for the further propagation of the crack would be that it follows the valley of compressive strain with tensile strain in its wake. Fig.~\ref{fig:crack_prediction}(b) shows the corresponding prediction.

\begin{figure}[htbp]
    \centering
    \includegraphics[width=1\textwidth]{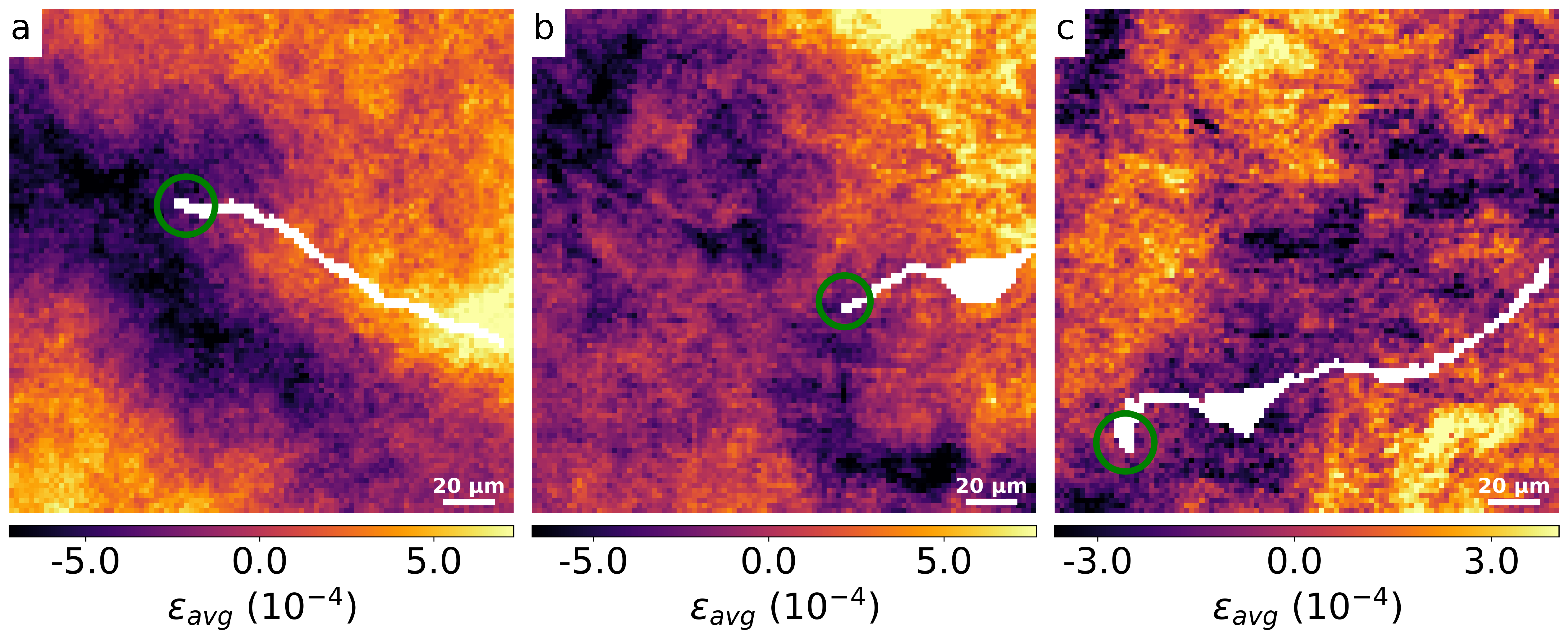}
    \caption{$\averagestrain$ maps from the crack tip for three different samples: (a) and (b) are shot-peened, while (c) is unpeened.}
    \label{fig:strain_cracktips}
\end{figure}

To determine the strain field around the crack tip rather than along the crack path, high-resolution scans with a step size of 1 micrometer was conducted in the crack-tip region.  Following the same procedure applied to the above scan, the resulting $\averagestrain$ maps for three distinct samples are depicted in Fig.~\ref{fig:strain_cracktips}(a-c). Fig.~\ref{fig:strain_cracktips}(a) corresponds to the sample previously analyzed in the overview scan and data analysis sections, which is shot-peened, while Fig.~\ref{fig:strain_cracktips}(b) and Fig.~\ref{fig:strain_cracktips}(c) illustrate the results for two additional samples, with (b) being shot-peened and (c) being unpeened. A comparable strain distribution around the crack tip in a fatigue-cracked Al-Li alloy sample was reported by Steuwer et al.~\cite{steuwer2010evolution}, showing tensile strain ahead of the crack tip and compressive strain in the wake. In contrast, our results indicate compressive strain ahead of the crack and tensile strain in the wake.

\section{Discussion} 

The initial analysis revealed effects of dynamic crystal recovery at the vicinity of the crack (see Fig.~\ref{fig:recrystallization_analysis}). This provided a basis for further investigation of the strain distribution along the surface. The $\averagestrain$ strain map from the initial overview scan shows strain distribution predominately along the surface (see Fig.~\ref{fig:epsilon}(d)). The subsequent high-resolution scan focused at the crack-tip offered a detailed view of the localized strain behavior. The strain field at the crack tip reveals a significant departure from the predictions of Linear Elastic Fracture Mechanics (LEFM). LEFM anticipates strain singularities at crack tips, where strain magnitudes theoretically approach infinity under idealized conditions~\cite{anderson2005fracture}. 

However, the $\averagestrain$ strain maps from the crack-tip region (see Fig.~\ref{fig:strain_cracktips}(a-c)) indicate that such singularities are mitigated in practice. Fig.~\ref{fig:strain_cracktips}(a), corresponding to the sample from the overview scan, and Fig.~\ref{fig:strain_cracktips}(b), representing another shot-peened sample, both show strain increasing along the crack path from the crack tip. In contrast, Fig.~\ref{fig:strain_cracktips}(c), representing the unpeened sample, displays no clear strain pattern. These observations indicate a deviation from LEFM expectations, highlighting the need for further investigation into the strain behavior at crack tips in martensitic steel.

\section{Conclusion}

In this study, we employed XRD scanning of martensitic steel samples, including an overview scan and targeted scans at the crack tip, to determine the strain field around the crack tip. By analyzing the integrated intensity, peak position, and peak width maps, we identified strain variations near the cracks. The strain  $\averagestrain$ at the crack tip was calculated, revealing patterns that deviate from the singularities predicted by Linear Elastic Fracture Mechanics. The $\averagestrain$ maps from three samples showed distinct behaviors: In the first two shot-peened samples, strain increases strain from crack tip along the crack path, while the third, unpeened sample, displayed no clear strain pattern. These results underscore the need for further research to understand the strain behavior at crack tips in martensitic steel.

%In the first two shot-peened samples, strain increases strain from crack ...". The second "strain" is too much, I think.

\begin{acknowledgements}

We acknowledge DESY (Hamburg, Germany), a member of the Helmholtz Association HGF, for the
provision of experimental facilities. Parts of this research were carried out at PETRA III, and we
would like to thank the staff for their assistance in using the P06 beamline. Beamtime was allocated
for proposal ID I-20220505. This research was supported in part through the Maxwell computational
resources operated at DESY.

\end{acknowledgements}

\begin{funding}

We thank the Federal Ministry of Education and Research (BMBF) for funding under Grant No. 05K22PS2 for the High Speed X-ray Nano Diffraction (HS-XRND) project.
\end{funding}

%\ConflictsOfInterest{Please declare any conflicts of interest, or declare  that there are no conflicts of interest.
%}

%\DataAvailability{Please state how the data supporting the results reported in your article can be accessed, e.g. within the article, as published supporting material, in repositories, upon request...
%}

\bibliography{iucr} % basename of .bib file

\end{document}